\documentclass[twocolumn,prl,showpacs,preprintnumbers,amsmath,noamssymb,letterpaper,floatfix, aps, 10pt]{revtex4-1}

\usepackage{calrsfs}
\let\degree\relax

\usepackage{graphicx}
\usepackage{xspace}
\usepackage[amssymb]{SIunits}

\newcommand{\NBo}{\ensuremath{\mathit{Bo}}\xspace}
\newcommand{\ac}{\ensuremath{\alpha_{\mathrm{c}}}\xspace}
\newcommand{\ham}{\ensuremath{\mathcal{H}}\xspace}

\usepackage{xcolor}

\date{\today}
\begin{document}

\title{Droplets on Inclined Plates: Local and Global Hysteresis of Pinned Capillary Surfaces}

\author{Michiel Musterd}
\author{Volkert van Steijn}
\author{Chris R. Kleijn}
\author{Michiel T. Kreutzer}
\email{m.t.kreutzer@tudelft.nl}
\affiliation{Department of Chemical Engineering, Delft University of Technology, Julianalaan 136, 2628 BL Delft, The Netherlands}
% 68.35.Np adhesion at solid surfaces and interfaces
% 68.08.Bc wetting in liquid-solid interfaces
% 68.03.Cd Surface Tension
% 47.55.D- liquid drops

\begin{abstract}
Local contact line pinning prevents droplets from rearranging to minimal global energy, and models for droplets without pinning cannot predict their shape. We show that experiments are much better described by a theory, developed herein, that does account for the constrained contact line motion, using as example droplets on tilted plates. We map out their shapes in suitable phase spaces. For 2D droplets, the critical point of maximum tilt depends on the hysteresis range and Bond number. In 3D, it also depends on the initial width, highlighting the importance of the deposition history.
\end{abstract}

\maketitle

\vspace{2cm}

The diverse and complex shapes of raindrops on a window strikingly illustrate the difficulty in understanding shapes of droplets under the influence of surface tension and gravity. Early theoretical work by Laplace, Young and Gauss \citep{Laplace1805,*Young1805,*Gauss1830} showed that at equilibrium, a droplet touches a solid surface at a unique angle, the \emph{Young contact angle} $\theta_Y$. In practice, however, the contact angle of static droplets often deviates from the Young angle, because the contact line gets pinned on physical or chemical defects before it has equilibrated to the lowest energy \citep{Johnson1964,*Joanny1984,*Quere2008,*DeGennes2003}. This results in a net force at the contact line, which can, akin to friction, balance gravity or shear in static droplets or slow down moving droplets. The range over which the angle can vary is bracketed by a receding angle $\theta_r$ and an advancing angle $\theta_a$, as has been observed for stationary droplets and moving droplets alike \citep{Rio2005,*snoeijer}. They depend on the density of surface defects \citep{Reyssat2009} and are often treated as constants for a given liquid-substrate combination, although it is observed and understood that these parameters are in fact asymptotes for vanishing defect size relative to droplet size \citep{Marmur1994}. Contact lines with angles in the hysteresis range $[\theta_r,\theta_a]$ do not move, and this explains qualitatively why droplets can remain stuck. These immobile drops are not only fascinating to observe; the minimal force to set them in motion is highly relevant technically, e.g. for condensers, pesticide spraying and water-repelling surfaces \citep{Miljkovic2013,*Bergeron2000,*Mannetje}. 

\begin{figure}[tb]
 \includegraphics[width=1.0\columnwidth]{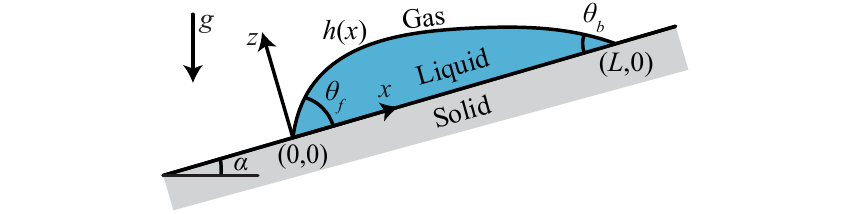}
\caption{Schematic 2D droplet in the coordinate system used for the Euler-Lagrange calculations 
}
\label{fig:1}
\end{figure}

A theory for droplet statics that takes this constrained contact line movement into account is still missing. Consider the classical experiment shown in Fig.~\ref{fig:1} that captures all the relevant physics: a sessile droplet on an inclined plate. Simplifications that have allowed theoretical progress in predicting the tilted droplet shape and the roll-off angle include fixing the contact line or the contact angle distribution along the contact line \citep{Dussan1983,*Carre1995}. However, these simplified geometries are at odds with experimental observations \citep{Berejnov2007}. Another approach has been to ignore the constraints entirely and analyze the problem as if the contact line is free to move \citep{Dimitrakopoulos1999,*Thampi2011}. In that equilibrium analysis, idealized models of sinusoidal microscopic roughness suggested that the roll-off angle corresponds to a much smaller hysteresis than found experimentally \citep{Krasovitski2005}, raising doubts about the validity of hysteresis ranges measured using tilting plates. Crucially, all of these approaches neither properly account for the constrained contact line movement, nor predict experiments accurately.

In this letter, we find droplet shapes by \emph{locally} taking the constrained movement (i.e. pinning when $\theta_r<\theta<\theta_a$) of the contact line into account. The crucial question is whether this simple constraint suffices to explain the rich features of the behavior of the entire droplet. It has been suggested \cite{Macdougall1942} that the entire droplet shape exhibits no hysteresis at all upon tilting back and forth, whereas we will show otherwise. We are interested in understanding how the local hysteresis of the contact line translates into hysteresis of the entire droplet shape and in critical behaviour of the transition from statics to dynamics. We begin our analysis for two-dimensional droplets, where significant analytical progress is possible, capturing most of the relevant phenomena, and then use numerical analysis of 3D droplets to compare with experiments. 

For a 2D droplet of volume $V$ in a reference frame as shown in Fig.~\ref{fig:1}, the effective interface Hamiltonian is given by \citep{Blossey2012}
\begin{align}
 \ham[h] =  E- p V = & \int_0^L dx \bigg[  \gamma \sqrt{1+(\partial_xh)^2} -\gamma \cos \theta_Y + \nonumber \bigg.\\  
& \left.+ \rho g \left(x h \sin \alpha +\frac{h^2}{2} \cos \alpha\right) - p h \right] 
\label{eq:minimizeE}
\end{align}
with $h(x)$ the shape of the gas-liquid interface. The first two terms under the integral are the surface energy of the gas-liquid and fluid-solid interfaces, with $\gamma$ the gas-liquid surface tension. The third term accounts for the potential energy for liquid density $\rho$ and tilt angle $\alpha$ and the last term is a Lagrange multiplier associated with fixed droplet volume that contains the Laplace pressure $p=\gamma \kappa$, where $\kappa$ is the mean curvature. The integral runs from $x = 0$ to $L$, the base length of the droplet. The first variation of \ham in dimensionless units then yields 
\begin{equation}
  \NBo\left(x \sin \alpha + h \cos \alpha\right) - \kappa - \frac{\partial_{xx} h}{\left[1+(\partial_x h)^2\right]^{3/2}} = 0
\label{eq:ELdroplet}
\end{equation}
where all lengths are in units of $V^{1/3}$ and the Bond number is defined as $\NBo=\rho g V^{2/3}/\gamma$. For given $\NBo$ and $\alpha$ the droplet shape $h(x)$ can be found by integrating this equation with appropriate boundary conditions. The first one, $h(0)=0$, fixes the coordinate system. The second one depends on the choice of the free parameters of the problem. In case $L$ is specified, then $h(L)=0$ completes the problem formulation and the contact angles at the front $ \theta_f=\tan^{-1} \partial_x h(0)$ and back $\theta_b=-\tan^{-1} \partial_x h(L)$ are a result of the calculation. Alternatively, if one of the angles is specified, $L$ follows from the calculation. After integration, $h(x)$ still contains the unknown parameter $\kappa$, which can be calculated using the volume constraint $\int h(x) dx=1$ and concludes the analysis. 

\begin{figure}[tb]
 \includegraphics[width=1.0\columnwidth]{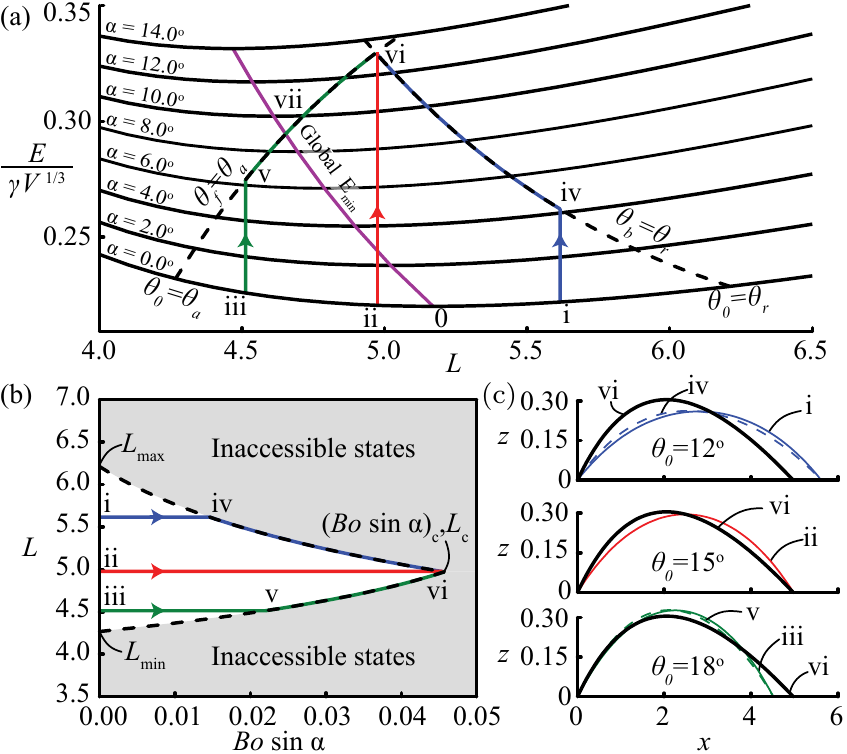}
\caption{(a) Energy phase space $E$-$L$ for a 2D droplet. (b) A phase diagram parameterized by observables $\NBo\sin\alpha$ and base length $L$. 
(c) Shape of three droplets with different initial contact angle, $\theta_0$, as they are deformed in the indicated parts of the tilt sequence (axes not to scale). Everywhere $\NBo=0.2$, $\theta_a=20\degree$, $\theta_r=10\degree$, $\theta_Y=14\degree$.}
\label{fig:2}
\end{figure}

We first map out all possible droplet shapes, $h(x)$, for given $\NBo$, $\alpha$ and $L$ without regarding the constraints set by the allowed contact angle range. To keep the analysis analytically tractable, we consider slender droplets for which ($\partial_x h)^2\ll 1$ in Eq.~(\ref{eq:ELdroplet}). The black lines in Fig.~\ref{fig:2}(a) show the dimensionless energy, $E/\gamma V^{1/3}$, corresponding to analytically calculated droplet shapes \footnote{For slender droplets, $h(x)=  b^{-2}\{\kappa - x \NBo \sin \alpha- [\kappa \sinh (bL-bx) + (\kappa- L \NBo\sin\alpha)\sinh bx ]/\sinh bL \}$ with $b=\sqrt{\NBo\cos\alpha}$ and $\kappa$ follows from $\int h(x) dx=1$.} parameterized by $L$ and $\alpha$ for $\NBo=0.2$. Indeed, the global energy minimum at $\alpha=0$ (point 0) corresponds to the shape of a sessile droplet that touches the substrate with the Young contact angle. For $\alpha>0$, the energy corresponding to these equilibrium shapes, that all have $\theta_b = \theta_Y$, can be calculated with a transversality condition \citep{Cassel} or constructed graphically by connecting the energy minima (purple line). Clearly, to remain at equilibrium, a droplet would have to be free to adapt its base length for any change in gravitational pull, but thermal fluctuations are too weak to facilitate this adaptation as long as the contact angles remain within the hysteresis range.

Before finding the actual evolution in the phase diagram in Fig.~\ref{fig:2}(a), we show how the hysteresis range $[\theta_r,\theta_a]$ puts constraints on the allowed values of $L$. It turns out that, for any $\alpha$, the shortest droplet is found by integrating Eq.~(\ref{eq:ELdroplet}) using $h(0)=0$ and $\partial_x h(0) = \tan\theta_a$, where the second root of $h(x)=0$ gives $L_\mathrm{min}$. Connecting the values of $L_\mathrm{min}$ at increasing $\alpha$ gives the $\theta_f=\theta_a$-curve in Fig.~\ref{fig:2}(a). Similarly, the $\theta_b=\theta_r$-curve for the longest droplet with base length $L_\mathrm{max}$ is found using $h(0)=0$ and $\partial_x h(L_\mathrm{max}) =- \tan\theta_r$. Including the constraints due to contact angle hysteresis hence reveals that the only permitted droplet shapes have an energy inside the area enclosed by the energy curve for $\alpha=0$ and the two curves corresponding to the smallest and largest $L$.

Knowing the permitted droplet shapes, we can describe the path taken by a droplet in the phase diagram. Consider the droplet that starts with an initial base length indicated by point (i). Upon tilting, the contact lines remain pinned ($\theta_r < \theta_b \leq \theta_f < \theta_a$) until point (iv) where $\theta_b = \theta_r$. Here the back depins and, tilting further, the droplet base shortens as it evolves to point (vi), where also the front depins: this is a unique \emph{critical} shape where $\theta_f = \theta_a$ and $\theta_b = \theta_r$ simultaneously. Below we describe how to calculate this critical value of $L$, here we mention that it depends on $\theta_a$ and $\theta_r$, not on the Young contact angle.

Two additional examples of full evolutions of droplets starting at points (ii) and (iii) are plotted in Fig.~\ref{fig:2}(a), and the actual evolution of $h(x)$ from initial shape to identical critical shape is shown in Fig. 2(c). Interestingly, all droplets tilted to the critical point and back to $\alpha = 0$ end up in point (ii). From there, they can be tilted back and forth between horizontal and critical point without shape hysteresis.  Although this behavior is more subtle for 3D droplets (see below), we have also observed it in experiments \citep{Musterd2013GFM}. 

This absence of a hysteresis loop implies no dissipation: indeed, the quasi-static deformation of the droplet analyzed here ignores viscous dissipation and a stationary contact line does not dissipate energy \citep{Priest2007}.

The energy diagram in Fig.~\ref{fig:2}(a) also resolves the debate raised by \citet{Krasovitski2005}. Our analysis does not assume periodic microscopic roughness and our results differ in details from \citep{Krasovitski2005}, but our work also demonstrates that equilibrium calculations predict an early roll-off with $\theta_f=\theta_a$ and $\theta_b=\theta_Y>\theta_r$ (point (vii) at $\simeq9\degree$). Yet, the analysis that does include pinning predicts roll-off later, with $\theta_f=\theta_a$ and $\theta_b=\theta_r$ at $\simeq13\degree$ (point vi). Of course, because contact lines in reality are pinned, $\theta_a$ and $\theta_r$ measured with tilting plates agree with the values measured by other means, as has been found experimentally \citep{Bormashenko2012} and now also explained theoretically.

A phase space that is more practical than the energy landscape in Fig.~\ref{fig:2}(a) is shown in Fig.~\ref{fig:2}(b), where all possible states are parameterized by $\NBo\sin\alpha$ and base length $L$. This phase space contains the same three examples as Fig.~\ref{fig:2}(a). The two boundaries are again given by shapes with $\theta_f = \theta_a$ and $\theta_b=\theta_r$, and all droplets trajectories meet eventually in the critical point $((\NBo\sin\alpha)_c,L_c)$ at the crossing of these boundaries. 
Using the analytically calculated shapes \footnote{At the critical point, the following expressions hold simultaneously: for the front $\partial_x h(0)=[2c^2/L^2+ b][2 c \coth (c/2)-4]^{-1}=\tan\theta_a$, and the back $\partial_x h(L)=[2c^2/L^2- b][2 c \coth (c/2)-4]^{-1}=-\tan\theta_r$. Here, $b=[c \coth (c/2)-2]^2 \tan\alpha$ and $c=L\sqrt{\NBo \cos\alpha}$. Solving for $L$ and $\NBo\sin\alpha$ yields $(\NBo\sin\alpha)_c=\left(\theta_a-\theta_r\right)\left(\theta_a+\theta_r\right)/2$}, one finds at the critical point 
$(\NBo\sin\alpha)_c=\left(\theta_a-\theta_r\right)\left(\theta_a+\theta_r\right)/2$, which is the slender drop approximation of the well known result $(\NBo\sin\alpha)_c=\cos\theta_r-\cos\theta_a$ for 2D droplets \citep{Macdougall1942}. 

We now approximate the phase space in Fig.~\ref{fig:2}(b) such that one is able to construct it without solving Eq.~(\ref{eq:ELdroplet}).  Approximating the boundaries with straight lines, this problem simplifies to finding expressions for $L_c$, $L_\mathrm{min}$ and $L_\mathrm{max}$ in addition to $(\NBo\sin\alpha)_c$ derived above. On a horizontal surface, relations for $L=f(\theta_0)$ are known for many situations \citep{Finn1988,*Shanahan1984,*Quere1998}: we, for example, find $L=(\theta_0/6-\NBo/60)^{-1/2}$ for 2D droplets up to $O(\NBo)$. Substitution of $\theta_0=\theta_a$ and $\theta_0=\theta_r$ readily gives $L_\mathrm{min}$ and $L_\mathrm{max}$. 
This leaves the critical value $L_c$. The red lines in Fig.~\ref{fig:2} represent the trajectory of a droplet that is initially at $\theta_0=(\theta_a+\theta_r)/2$ and suggest that its base length remains fixed up to the critical point \footnote{For slender droplets, the change of $\theta_f$ and $\theta_b$ with tilt is almost antisymmetric: $\partial_\alpha \theta_f/\partial_\alpha \theta_b=-1+\alpha (2\theta_0/5-\NBo/25)$ up to $O(Bo)$ and $O(\alpha)$. A droplet  that initially has $\theta_0=(\theta_a+\theta_r)/2$ hence advances to the critical point with negligible change in $L$. For non-slender droplets, we find the same conclusion by numerical integration of Eq.~(\ref{eq:ELdroplet}).}. Then, $L_c$ is found by calculating the length of a horizontal droplet of $\theta_0 = (\theta_a+\theta_r)/2$, which concludes the approximate calculation of all points in the phase diagram for 2D droplets.

\begin{figure}[tb]
 \includegraphics[width=1.0\columnwidth]{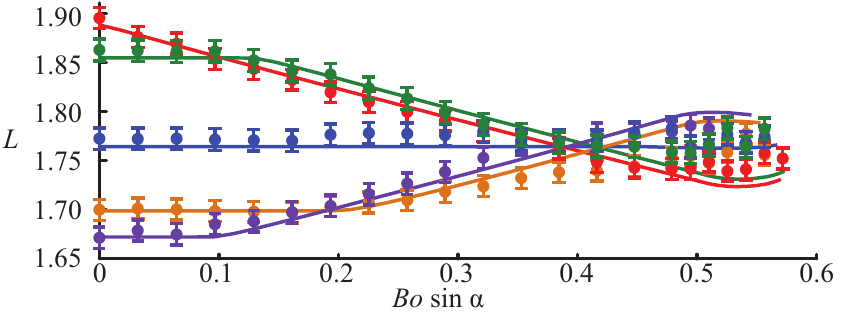}
\caption{Comparison of the experimental and numerical evolution of $L$ vs. dimensionless gravitational pull $\NBo\sin\alpha$ for droplets with initially circular base ($W=L$ at $\NBo \sin \alpha=0$). $\NBo=1.85$, $\theta_a=93\degree$, $\theta_r=74\degree$.}
\label{fig:3}
\end{figure}

We extend our analysis to 3D droplets using the code Surface Evolver \citep{Brakke1992}, adapted as in \citep{Santos2012,*Semprebon2014} to implement the local contact-line physics that we have also used in 2D. In 3D, the set of possible initial contact lines $h(x,y)=0$ is much larger than for 2D droplets (uniquely defined by $L$). After initializing such a base, we calculate the steady shape that minimizes the 3D equivalent of Eq.~(\ref{eq:minimizeE}) at increasing values of $\alpha$, using the solution at the previous tilt angle as initial condition. 
The resulting drop shapes have constant curvature, i.e. $\kappa-\NBo(z\cos\alpha+x\sin\alpha)$ is constant on the surface such that the fluid is at rest, and the local dimensionless pinning force $(\cos\theta-\cos\theta_Y)$ has a negative minimum at the front, a positive maximum at the back and passes through zero in between such that there always is a region that does not depin. 
Finally, we find the critical tilt angle, $\ac$, as the first value of $\alpha$ for which the droplet moves at each iteration, indicating roll-off. 

We have validated our simulations by experimentally measuring the phase diagram of $L$ vs. $\NBo\sin \alpha$. Deionized water droplets of given volume were positioned on a perfluorosilane coated silicon substrate on an automated tilting plate. The plate is enclosed in a chamber at 100\% humidity and tilted until roll-off in steps of $0.5-1^\circ$ with 60\,s equilibration time per tilt angle. Droplets with different initial circular bases were created using the well-known hysteresis loop in filling/emptying a droplet on a hysteretic surface \citep{RuizCabello2011}. 
Fig.~\ref{fig:3} shows the evolution of $L$ until roll-off for different initial conditions and the good agreement of the numerical results with the experiments. The value of $\cos\theta_r-\cos\theta_a$ is much higher than for the slender drops of Fig.~\ref{fig:2}, with $(Bo\sin\alpha)_c\approx0.55$. Interestingly, this critical point in 3D is not unique but depends on the initial shape.

\begin{figure}[tb]
 \includegraphics[width=1.0\columnwidth]{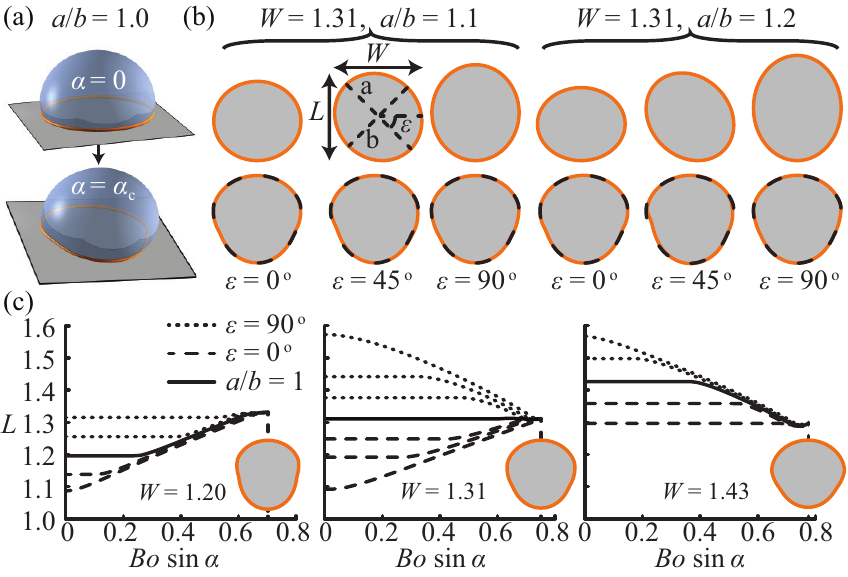}
\caption{(a) Rendering of simulated droplet shapes at $\alpha=0$ and $\alpha=\ac$ for $a/b=1$ (circular base)   (b) Base shapes for droplets with different initial elliptical shape, all having $W=1.31$. The overlayed roll-off shape for $a/b=1$ (dashed line) is added to illustrate the agreement in critical shape. (c) Numerically calculated evolution of the base length for droplets with elliptical base at three initial widths $W$. Everywhere $\NBo=1.5$, $\theta_a=140\degree$, $\theta_r=100\degree$.}
\label{fig:4}
\end{figure}

The simulations allow us to explore the influence of this initial shape of the contact line $h(x,y)=0$. We focus on elliptic shapes characterized by principle axes $a$ and $b$ and orientation $\epsilon$ (Fig.~\ref{fig:4}(b)) as a representative and experimentally realistic subset of all initial shapes.  Remarkably, all droplets of identical $\NBo, \theta_a$ and $\theta_r$ converge to the same critical point $((\NBo\sin\alpha)_c,L_c)$ if their initial width $W=a\cos\epsilon$ is the same, whatever the initial shape was, see Fig.~\ref{fig:4}(b) and the middle panel of Fig.~\ref{fig:4}(c). This finding is analogous to the 2D result that the critical droplet shape does not depend on the initial length. By contrast, we find different critical shapes for different initial values of $W$, and thus different critical points, shown in the other panels of Fig.~\ref{fig:4}(c). This finding stresses the importance of considering the constrained movement of the contact line:  attempts to find the critical droplet shape in 3D without taking the deformation history into account are doomed to fail.

\begin{figure}[tb]
 \includegraphics[width=\columnwidth]{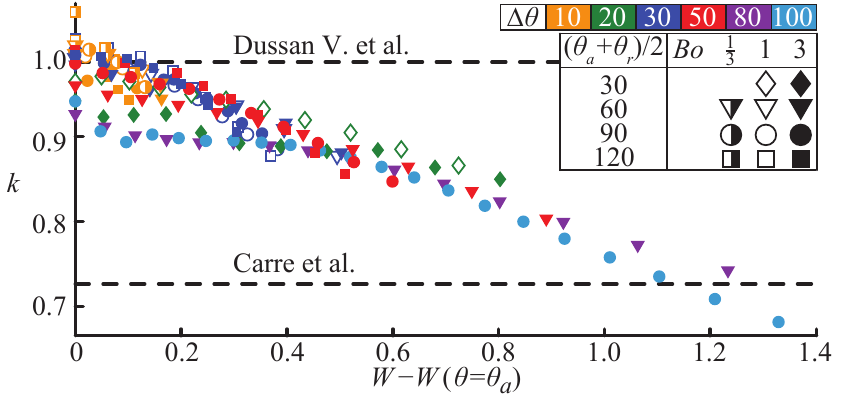}
\caption{Numerically calculated values of $k$ (Eq.~\ref{eq:k}) versus the base width $W$ for a wide range of $\Delta\theta=\theta_a-\theta_r$ (indicated by color), $\NBo$ and $(\theta_a+\theta_r)/2$ (indicated by marker shape and fill).  $W$ is shifted by $W(\theta=\theta_a)$, the base width of a droplet at the advancing contact angle, to highlight similar trends in the datasets.}
\label{fig:5}
\end{figure}

Critical points derived earlier \citep{Dussan1983,Furmidge1962} can be formulated as
\begin{equation}
(\NBo\sin\alpha)_c=k W_c(\cos\theta_r-\cos\theta_a), \label{eq:k}
\end{equation} 
as for 2D above, but now including the width $W_c$ of the critical shape and $k$, an $O(1)$ constant.
A first problem that limits the predictive power of Eq.~\ref{eq:k} is the unknown $W_c$.

We resolve this by noting that in the course of tilting, the width does not change: there is no force to move the contact line perpendicular to the direction of gravity and the critical width $W_c$ equals the initial width. 
A second problem is that contact line shapes assumed in earlier work \citep{Dussan1983}, i.e. circles or curves connected by straight segments parallel with gravity, are at odds with experimental observations \citep{Berejnov2007} and our simulated critical shapes (Fig.~\ref{fig:4}). As the precise value of $k$ depends on details of the critical shape, theoretical progress will amount to predicting the full evolution from initial to critical shape, where one has the freedom to pick the most convenient initial shape of a given width. The values calculated with our simulations are summarized in Fig.~\ref{fig:5}, together with the theoretical predictions $k=1$ and $k=\pi/4$ \citep{Carre1995}. Clearly, the present analysis shows that $k$ is not a constant, and the most prominent trend is that $k$ decreases with increasing initial width, from values close to $k=1$ for the smallest possible width for a given $\theta_a$. 

In summary, we have shown that capillary surfaces as observed in experiments can only be calculated by considering the full evolution from initial conditions, because the constrained movement of pinned contact lines prohibits the bodies enclosed by such capillary surfaces from sampling the entire phase space. As simplest yet complete test case, we have considered sessile droplets on tilted plates.  Including hysteretic behavior locally at the contact line properly describes the evolution of droplets, and teaches to what extent global critical behavior depends on this local hysteresis. For 2D droplets, both contact angles depin at one unique state, whereas in 3D droplets parts of the contact line lack a driving force to trigger depinning, such that initial hysteresis remains relevant even at the critical state. As a result, progress in predicting when droplets succomb to pull is only possible if the deposition history is known.

We wish to thank K.A. Brakke for valuable input on the Surface Evolver simulations. This research was carried out in the framework of the HESTRE project of ISPT.

\end{document}